\documentclass[12pt]{article}
\usepackage{graphicx}
\usepackage{amsmath}
\usepackage{bm}
\def\Vec#1{\mbox{\boldmath $#1$}}

\def\itmb{\begin{itemize}}
\def\itme{\end{itemize}}
\def\enmb{\begin{enumerate}}
\def\enme{\end{enumerate}}
\def\eqnb{\begin{equation}}
\def\eqne{\end{equation}}

\def\NPB{{Nucl. Phys.} { B}}
\def\PLB{{Phys. Lett.} B}
\def\PRL{Phys. Rev. Lett.}
\def\PRD{{Phys. Rev.} D}
\def\PRC{{Phys. Rev.} C}

\def\RMP{{Rev. Mod. Phys.}}

\title{Axial anomaly and the triality symmetry of octonion}

\author{Sadataka Furui \\
  Faculty of Science and Engineering, Teikyo University\\
1-1 Toyosatodai, Utsunomiya, 320-8551 Japan {\thanks
{\textit{E-mail address:} furui@umb.teikyo-u.ac.jp}}
}

\begin{document}
\maketitle
\begin{abstract}%
With an assumption that in the Yang-Mills Lagrangian,  a left-handed fermion and a right-handed fermion both expressed as the quaternion makes an octonion, and the gauge field can be treated as self-dual,  we calculate the  axial current and two vector currents triangle diagram of Bardeen, which yields the contribution of the axial anomaly.

The octonion possesses the triality symmetry, and there are 5 symmetry operations $G_{ij}$ and $G_{ijk}$  ($ijk=123)$, in which mixing of spinors and vectors occur.  $G_{23}$ does not mix vectors and spinors, but mismatch of the spinor and vector fields occurs. Hence, electro magnetic (EM) wave emitted from galaxies transformed by the five transformations would not be detected by EM detectors in our galaxy, and the source would be regarded as dark matter. 

 The axial anomaly appears as a reflection of the symmetry of the matter field and not as a reflection of the symmetry of the pure vacuum, which is consistent with recent arguments on condensates and confinement. 
\end{abstract}

\newpage
\section{Introduction}
In the infrared field theory, instanton or topological effect in the QCD is important. 
The QCD running coupling constant $a(Q^2)=\alpha_s(Q^2)/\pi$ at large $Q^2$ is parametrized as
\[
a(Q^2)\sim \frac{\beta_0}{\log(Q^2/\Lambda^2)}
\]
where $\displaystyle \beta_0=\frac{1}{4}(11-\frac{2}{3}N_f)$ and $\Lambda$ is the QCD scale.
The QCD $\beta$ function is defined as
\[
\beta(a)=Q^2\frac{da}{dQ^2}=-\beta_0 a^2-\beta_1 a^3+\cdots=-\beta_0 a^2(1+c a+c a^2+\cdots)
\]
At $N_f=16.5$, $\beta_0$ becomes 0, and at $N_f\sim 8$, $\displaystyle c=\frac{\beta_1}{\beta_0}=\frac{1}{4\beta_0}[102-\frac{38}{3}N_f]$ becomes 0.

At low energy, perturbative approach
\[
a_R(Q^2)=a(\mu^2)+r_1 a(\mu^2)^2+r_2 a(\mu^2)^3+\cdots
\]
 is not promising, and Brodsky, Lepage and Mackenzie(BLM)\cite{BLM83} introduced a scale dependent expansion scheme
\[
a_R(Q^2)=a(k_0^2)+c_1 a(k_1^2)^2+c_2 a(k_2^2)^3+\cdots
\]
where $k_i^2$ are different scales proportional to the external scale $Q^2$. Coefficients $a(k^2)$ are assumed to satisfy the renormalization equation
\[
\frac{d a(k^2)}{d\log k^2}=-(\beta_0 a(k^2)^2 +\beta_1 a(k^2)^3+\beta_2 a(k^2)^4+\cdots)\def \overline{\beta(a)}
\]

In the region where $c_1$ is close to zero, a conformal window could open and the infrared fixed point that controls physics in this energy region could appear.
BLM considered the region where $c_1$ is small, and incorporated $N_f$ dependence in the choice of $k_0^2, k_1^2$ and $k_3^2$ etc. and  $\beta_i$ are taken to be $N_f$ independent.  Grunberg\cite{Gr89} also defined expansion of $\overline{\beta(a)}$ appropriate for the low-energy physics, which is called effective charge method(ECH), and Gardi et al.\cite{GGK98} showed that presence of infrared fixed point is possible when $N_f>5$.  Relations between observables and the infrared fixed point and scheme dependence are investigated also in \cite{GK98}, and freezing for $N_F\geq 5$ was suggested. 

Banks and Zaks(BZ)\cite{BZ82} solved the equation $\beta(a)=0$ for some $N_f$ such that $\beta_0$ is small and $\beta_1$ is negative and defined $a_{FP}=-\beta_0/\beta_1$.  With the expansion parameter
\begin{equation}
a_0= -\frac{\beta_0}{\beta_1|_{\beta_0=0}}=\frac{\beta_0}{-\beta_{1,0}}
\end{equation}
$a_R(Q^2)$ is expanded as polynomials of $a_0^n$.
Based on this perturbative expansion, Appelquist et al.\cite{AFN08} performed a lattice simulation of QCD running coupling  in the Schroedinger functional(SF) scheme. They found that large numbers of flavors are necessary for making the running coupling in the infrared insensitive to the momentum squared.
They concluded that for a presence of an infrared fixed point $N_f\sim 12$ is necessary.

The conformal window of QCD ${N_f}^{c}\simeq 8$ was claimes also in the staggered fermion lattice simulation\cite{CHPS13} and in full QCD lattice simulation\cite{BRRQ13}. In the case of staggered fermion, spontaneous simmetry breaking of single-site shift symmetry $S^4$\cite{SCHP12} was considered and in the full QCD lattice simulation, Landau gauge fixing was adopted\cite{ABBCRQ12}.
 
 Brodsky et al.\cite{BGGR01} investigated the relation between the conformal effect and the infrared fixed point of QCD.  The renormalization scheme of BLM\cite{BLM83} is based on the MOM scheme, and measurable quantities $\rho$ is expanded in powers of effective running coupling as 
\begin{equation}
\rho=C_0\alpha_s(Q^*)[1+C_1^*\frac{\alpha_s(Q^*)}{\pi}+\cdots]
\end{equation}
where $C_1^*$ and $Q^*$ are $N_F$ independent.

 A lattice simulation of QCD running coupling in the momentum subtraction scheme(MOM)\cite{FN08} suggests that the system of $N_F=2+1$ is not far from the conformal window. 
The critical flavor number of QCD for a presence of the conformal window calculated in the Banks-Zaks expansion was ${4\leq N_f^*\leq 6}$ and $N_f/N_c=3/2$ \cite{Gr02}. The electron scattering data \cite{BTD10} also suggests proximity of the infrared fixed point.
In the SF scheme, matching of coordinate space wave function in different scales is performed, and the large flavor number $N_f$ necessary for the presence of a conformal window.
In \cite{SF12}, we investigated the symmetry of a four-component spinor which is transformed by an octonion, which is a combination of two quaternions.  An octonion has the triality symmetry and the matching of in the SF scheme could be the matching of both flavor and the triality.

In the nature, there could be a massless gauge boson which is a combination of a photon and a gluon, but electron sees the photon part only\cite{Wilczek11}. Photons and gluons could have qualitative differences. 
In QED, we assume that the electro magnetic interaction breaks the triality symmetry, and leptons interacts with photons or gamma rays in the same triality sector.  Photons or gamma lays in different triality sector than that of electrons in the detector will be regarded as the dark matter.

When the triality symmetry of leptons are broken, one can produce two degenerate light leptons and one heavy lepton, which belong to a different triality sectors\cite{SF12}. In the strong interaction, no triality selection rules are expected, and the strong interaction of hadrons will not be modified.

The triality symmetry and the number of flavors is important also in neutrino oscillation in reactor and in astrophysics.  In the analysis of reactor experiment data of neutrino oscillation, 3+2 flavor model\cite{MMPPSS08} was successful. 
In order to explain the diffuse $1\sim 10$MeV gamma ray emission and the 511 keV line flux from
the Galactic bulge\cite{YK08}, a decay of weakly interacting massive particle (WIMP) dark matter is proposed\cite{CS08}.  
As a candidate of the WIMP dark matter, a brane world model with large extra dimensions, massive neutrinos etc. are proposed.  Recent observation of the shape of Milky way halo suggests that the dark matter is warm, and its mass is around 10keV. The result excludes sterile neutrino as the WIMP. 

Neutrinos and leptons are represented by spinors. The transformation of a two-component spinor is  expressed by a quaternion, and a four component spinor whose transformation is expressed by a sum of two quaternions can transform as an octonion, which has the triality symmetry.
The neutrinos in a different triality sector could have different mass. Neutrino in a different triality sector than that of corresponding lepton would become the warm dark matter.

\section{Triality symmetry of octonion}
The four component Dirac spinor is a combination of two two-component spinors each transforms by quaternions.
\'E. Cartan\cite{Cartan66} defined, using semi-spinors of an even number of indices
\[
   \xi_{even}:= \xi_{12},\xi_{23}, \xi_{34}, \xi_{13}, \xi_{24}, \xi_{14}, \xi_{1234}, \xi_0
\]
and semi-spinors of an odd number of indices
\[
   \xi_{odd}:= \xi_{1}, \xi_{2}, \xi_{3}, \xi_{4}, \xi_{234}, \xi_{134}, \xi_{124}, \xi_{123},
\]
four bases of spinors 
\begin{eqnarray}
 &&A= \xi_{14}\sigma_x+\xi_{24}\sigma_y+\xi_{34}\sigma_z+\xi_{0} {\bf I}\nonumber\\
 &&B= \xi_{23}\sigma_x+\xi_{31}\sigma_y+\xi_{12}\sigma_z+\xi_{1234} {\bf I}\nonumber\\
 &&C= \xi_{1}\sigma_x+\xi_{2}\sigma_y+\xi_{3}\sigma_z+\xi_{4} {\bf I}\nonumber\\
 &&D= \xi_{234}\sigma_x+\xi_{314}\sigma_y+\xi_{124}\sigma_z+\xi_{123} {\bf I}
\end{eqnarray}
and two bases of vectors
\begin{eqnarray}
&&E=x_1 {\Vec i}+x_2{\Vec j}+x_3{\Vec k}+x_4{\bf I}\nonumber\\
&&E'=x_1' {\Vec i}+x_2'{\Vec j}+x_3'{\Vec k}+x_4'{\bf I}.
\end{eqnarray}

He studied a superspace transformations $G$, which makes the trilinear form 
\begin{eqnarray}
{\mathcal F}&&={^t\phi} CX\psi=x^1(\xi_{12}\xi_{314}-\xi_{31}\xi_{124}-\xi_{14}\xi_{123}+\xi_{1234}\xi_1)\nonumber\\
&&+x^2(\xi_{23}\xi_{124}-\xi_{12}\xi_{234}-\xi_{24}\xi_{123}+\xi_{1234}\xi_2)\nonumber\\
&&+x^3(\xi_{31}\xi_{234}-\xi_{23}\xi_{314}-\xi_{34}\xi_{123}+\xi_{1234}\xi_3)\nonumber\\
&&+x^4(-\xi_{14}\xi_{234}-\xi_{24}\xi_{314}-\xi_{34}\xi_{124}+\xi_{1234}\xi_4)\nonumber\\
&&+x^{1'}(-\xi_{0}\xi_{234}+\xi_{23}\xi_{4}-\xi_{24}\xi_{3}+\xi_{34}\xi_2)\nonumber\\
&&+x^{2'}(-\xi_{0}\xi_{314}+\xi_{31}\xi_{4}-\xi_{34}\xi_{1}+\xi_{14}\xi_3)\nonumber\\
&&+x^{3'}(-\xi_{0}\xi_{124}+\xi_{12}\xi_{4}-\xi_{14}\xi_{2}+\xi_{24}\xi_1)\nonumber\\
&&+x^{4'}(\xi_{0}\xi_{123}-\xi_{23}\xi_{1}-\xi_{31}\xi_{2}-\xi_{12}\xi_3)\nonumber
\end{eqnarray}
and scalar products
\begin{eqnarray}
F&&=x_1 x_1'+x_2 x_2'+x_3 x_3'+x_4 x_4'\nonumber\\
\Phi&&=\xi_0\xi_{1234}-\xi_{23}\xi_{14}-\xi_{31}\xi_{24}-\xi_{12}\xi_{34}\nonumber\\
\Psi&&=-\xi_1\xi_{234}-\xi_2\xi_{314}-\xi_3\xi_{124}+\xi_4\xi_{123}\nonumber\\
\end{eqnarray}
invariant.

Dirac Fermions are expressed as $\eta=\left(\begin{array}{c} A\\
                                            B\end{array}\right)$ and 
$\chi=\left(\begin{array}{c} C\\
                              D\end{array}\right)$.

There are 5 superspace transformations $G_{23}, G_{12}, G_{13}, G_{123}$ and $G_{132}$, whose operation on spinors and vectors are given in Tab.1 and 2.

\begin{tabular}{lll}
$G_{23}$& $G_{12}$& $G_{13}$\\
\hline
  $A\to (C_1,C_2,C_3,C_4)$ &  $A\to (x_1', x_2', x_3', x_4)$ & $A\to (A_1,A_2,A_3,B_4)$ \\
  $B\to (D_1,D_2,D_3,D_4)$ & $B\to (x_1,x_2,x_3,x_4')$ & $B\to (B_1,B_2,B_3,A_4)$\\
  $C\to (A_1,A_2,A_3,A_4)$ & $C\to (C_1,C_2,C_3,D_4)$ &  $C\to (x_1', x_2', x_3', x_4)$\\
  $D\to (B_1,B_2,B_3,B_4)$ & $D\to (D_1,D_2,D_3,C_4)$ & $D\to (x_1,x_2,x_3,x_4')$\\
   $E\to (x_1,x_2,x_3,x_4')$ & $E\to (B_1,B_2,B_3,A_4)$ & $E\to (D_1,D_2,D_3,D_4)$ \\
    $E'\to (x_1',x_2', x_3', x_4)$ & $E'\to (A_1,A_2,A_3,B_4)$ & $E'\to (C_1,C_2,C_3,C_4)$\\
\hline

\label{Tab. 1 }
\end{tabular}

\begin{tabular}{ll}
$G_{123}$& $G_{132}$\\
\hline
 $A\to (x_1', x_2', x_3', x_4')$ &  $A\to (C_1,C_2,C_3,D_4)$ \\
  $B\to (x_1,x_2,x_3,x_4)$ &  $B\to (D_1,D_2,D_3,C_4)$\\
 $C\to (A_1,A_2,A_3,B_4)$ & $C\to (x_1',x_2',x_3,'x_4)$ \\
$D\to (B_1,B_2,B_3,A_4)$ &  $D\to (x_1,x_2,x_3,x_4')$ \\
 $E\to (D_1,D_2,D_3,C_4)$ & $E\to (B_1,B_2,B_3,B_4)$ \\
 $E'\to (C_1,C_2,C_3,D_4)$ & $E'\to (A_1,A_2,A_3,A_4)$ \\
\hline\label{Tab. 2}
\end{tabular}

Except for $G_{23}$, transformation of spinors to vectors and vectors to spinors occur.  I consider the invariant trilinear form in the coupling of fermions and vector particles.
 On $S^4=R^4\cup\{\infty\}$, one can define stereographic projections from the north pole
$F_N:S^4-\{N=(0,0,0,1)\}$ and $F_S:S^4-\{S=(0,0,0,-1)\}$ 
 Cover  $S^4$ by two hemispheres $U_+$ and $U_-$ with the overlap $U_+\cap U_-=S^3\times [-\epsilon,\epsilon]$ and one can define the instanton. 

 The Dirac fermions ${^t(}\phi,c\phi,\psi, c\psi )$ is transformed by $G_{23}$ to Dirac fermions whose second pair is the complex conjugate of the first pair multiplied by -1. 
\[
G_{23}\left(\begin{array}{c} \phi\\
                             c\phi\\
                              \psi\\
                              c\psi\end{array}\right)=
 \left(\begin{array}{c} \psi\\
                       c\psi\\
                        -\phi^*\\
                        -c\phi^*\end{array}\right)
\]
 We put $\psi\sim e^{-ikx+iE t}$ on the $U_+$ and $-\phi^*\sim -e^{-ikx-iEt}$ on the $U_-$ and assume the transformed $U_+\cap U_-$ appears as an instanton in our world and the rest appear as the dark matter due to mismatch of phases.

\section{Axial anomaly and triality symmetry}
In 1969, Bardeen\cite{Ba69} showed that the vector and axial vector currents associated with
the external vector and axial vector fields satisfies anomalous ward identities, and derived the
explicit logarithmically divergent four-vertex loop  contribution. 
In the Standard Model, EM anomaly of the axial isovector current\cite{ZWZ83}
\[
\partial _\mu j^{\mu 5 3}=-\frac{e^2}{16\pi^2}\epsilon^{\alpha\beta\mu\nu}F_{\alpha\beta}F_{\mu\nu} \frac{1}{2}
\]
where superfix 3 is the index in the isospin space.
The non-abelian chiral anomaly is derived from the $D=4$ field lagrangian\cite{ZWZ83} 
\begin{equation}
{\mathcal L}=\bar \psi_L i\gamma^\mu (\partial_\mu-iA^i_{L\mu}\lambda_i)\psi_L
+\bar\psi_R i\gamma^\mu(\partial_\mu-i A^i_{R\mu}\lambda_i)\psi_R
\end{equation}
where $A^i_\mu={\mathcal V}^i_\mu+{\mathcal A}^i_\mu\gamma_5$.

The quantum action functional $W(A)$ expressed by
\[
e^{iW(A_\mu)}=\int{\mathcal D}\psi{\mathcal D}\bar\psi e^{i\int {\mathcal L}d^4 x}
\]
satisfies under the transformation
\[
\psi(x)\to e^{i\theta^i(x)\lambda_i\gamma_5}\psi(x)
\]
the anomalous Ward identity
\[
\frac{\delta W[A^\theta_\mu]}{\delta \theta^i(x)}|_{\theta^i(x)=0}=G_i(x)
\]
According to Wess and Zumino \cite{WZ71}, the identity is expressed as
\[
\partial _\mu\frac{\delta}{\delta A_{\mu_i}}+(A_\mu\times \frac{\delta}{\delta A_\mu})_i W=G_i(A)
\] 
The anomaly under $SU(3)\times SU(3)$ symmetry has the form
\begin{equation}
D^\mu J_{\mu i}=-\frac{1}{24\pi^2} tr [\lambda^i \partial^\mu \epsilon_{\mu\nu\rho\sigma}(A^\nu\partial^\rho A^\sigma+\frac{1}{2}A^\nu A^\rho A^\sigma)].
\end{equation}
I consider gauge configurations in Coulomb gauge and consider
\begin{equation}
D^0 J_{0 i}=-\frac{1}{24\pi^2} tr [\lambda^i \partial^0 \epsilon_{0 k {\l} m}(A^k{\partial^{\l}} A^m+\frac{1}{2}A^k A^{\l} A^m ) ]
\end{equation}
where $k,\l,m=1,2,3$.

The coupling of the anomaly to $J_0^{5a}=\bar\psi T^a \gamma_0\gamma_5\psi$ is, 
\begin{eqnarray}
&&D^0 J_{0}^5=-\frac{1}{4\pi^2}\epsilon^{0\nu\rho\sigma}Tr[\{T^a[\frac{1}{4}V_{4\nu}V_{\rho\sigma}+\frac{1}{12}A_{4\nu}A_{\rho\sigma}\nonumber\\
&&+\frac{2i}{3}(A_0 A_\nu V_{\rho\sigma}+A_0 V_{\nu\rho}A_\sigma+V_{0\nu}A_\rho A_\sigma)\nonumber\\
&&-\frac{8}{3}A_0 A_\nu A_\rho A_\sigma]\}]
\end{eqnarray}
where 
\[
V_{\mu\nu}=\partial_\mu V_\nu-\partial_\nu V_\mu-i[V_\mu, V_\nu]-i[A_\mu, A_\nu]
\]
and
\[
A_{\mu\nu}=\partial_\mu A_\nu-\partial_\nu A_\mu-i[V_\mu, A_\nu]-i[A_\mu, V_\nu]
\]

As the vertex renormalization of the triangle diagram of the axial vector - vector - vector vertex, the photon rescattering diagram was considered\cite{AI89, Adler04, Ioffe08}. 
\begin{figure}
\begin{center}
\includegraphics[width=6cm,angle=0,clip]{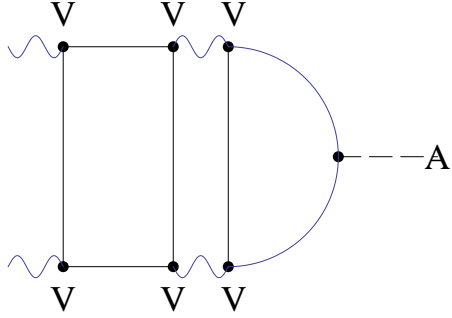}
\caption{The axial anomaly vertex diagram with final state vector field-lepton rescattering.}
\label{g3}
\end{center}
\end{figure}
The divergence of the axial vector current that couples to the topological anomaly reads
\begin{equation}
\partial^\mu{\mathcal F}^5_\mu(x)=2i m_0 j^5(x)+\frac{\alpha_0}{4}F^{\xi\sigma}(x)F^{\tau\rho}(x)\epsilon_{\xi\sigma\tau\rho} \label{divergence}
\end{equation}

Renormalization of the axial current anomaly including the rescattering diagram (Fig.1) was investigated in\cite{AI89}. 
It was shown by \cite{Adler04} that the axial current vertex renormalization factor $Z_A$ and the wave function renormalization factor $Z_2$ are related by
\begin{equation}
Z_A=Z_2[1+\frac{3}{4}(\alpha_0/\pi)^2 \log(\Lambda^2/\mu^2)+\cdots]
\end{equation}
and
\[
\langle \partial^\mu j_\mu^5\rangle=\langle F_{\mu\nu}\hat F^{\mu \nu} \rangle_{ext} (1-\frac{3 e_0^4}{64\pi^2}\log\frac{\Lambda^2}{m^2})
\]

Using the dual quaternion bases $\psi$, $\phi$  and their charge conjugate partner $C\psi, C\phi$, it is possible to make three vector particle exchange diagrams that introduce the axial anomalies as shown in Figs.2-13.  
We adopt the Coulomb gauge and the gauge particle that couples to $x_4$ or $x_4'$ satisfies $\vec \nabla\cdot \vec A=0$\cite{BZ99, BZ07}.  

\begin{figure}
\begin{minipage}[b]{0.47\linewidth}
\begin{center}
\includegraphics[width=6cm,angle=0,clip]{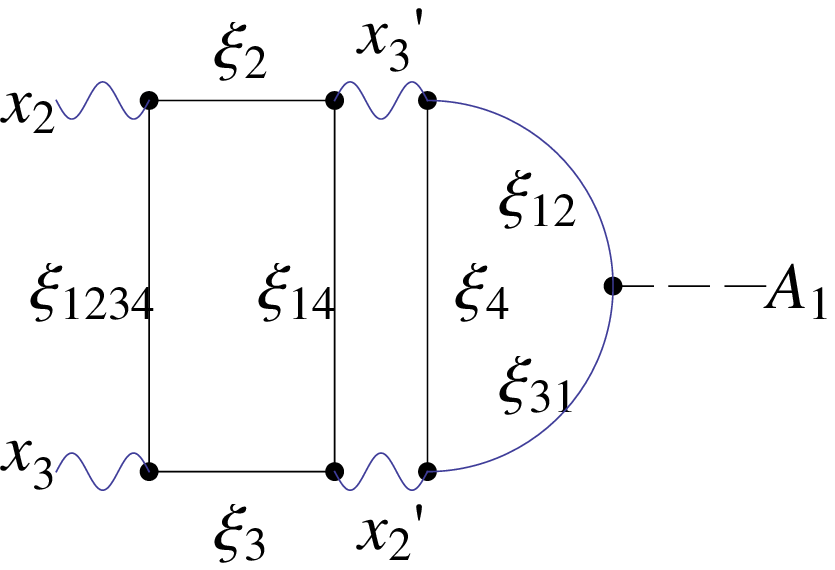}
\caption{The triangle loop diagram of axial anomaly. $A_1x_3' x_2'$ type and its rescattering.} 
\label{g1a}
\end{center}
\end{minipage}
\hfill
\begin{minipage}[b]{0.47\linewidth}
\begin{center}
\includegraphics[width=6cm,angle=0,clip]{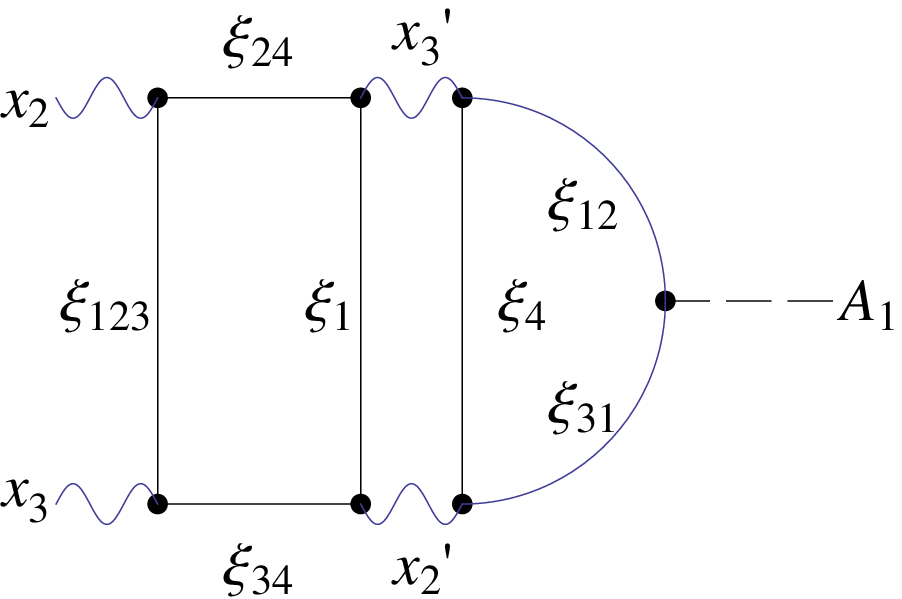}
\caption{The triangle loop diagram of axial anomaly. $A_1 x_3' x_2'$ type and its rescattering.}
\label{g1e}
\end{center}
\end{minipage}
\end{figure}
\begin{figure}
\begin{minipage}[b]{0.47\linewidth}
\begin{center}
\includegraphics[width=6cm,angle=0,clip]{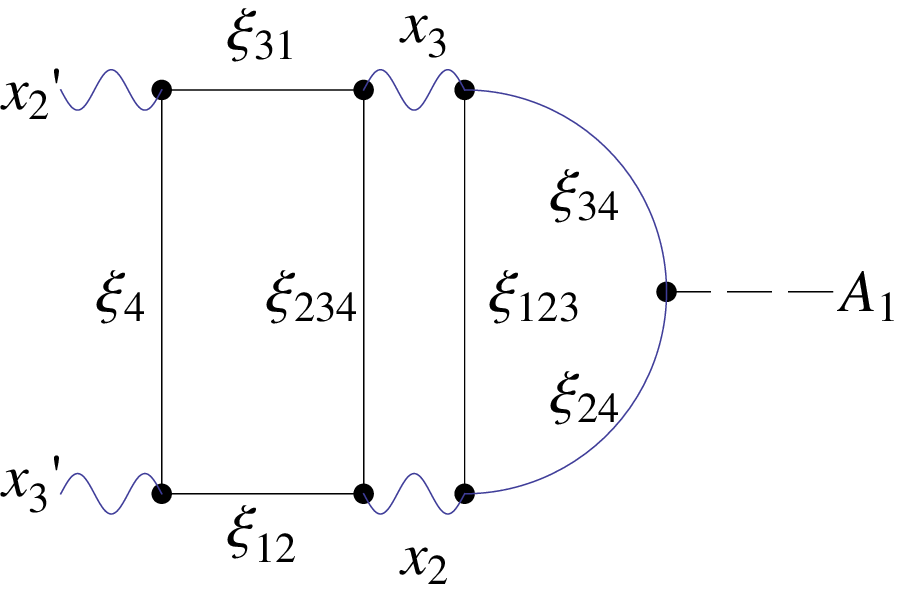}
\caption{The triangle loop diagram of axial anomaly. $A_1 x_3' x_2'$ type and its rescattering.} 
\label{g1c}
\end{center}
\end{minipage}
\hfill
\begin{minipage}[b]{0.47\linewidth}
\begin{center}
\includegraphics[width=6cm,angle=0,clip]{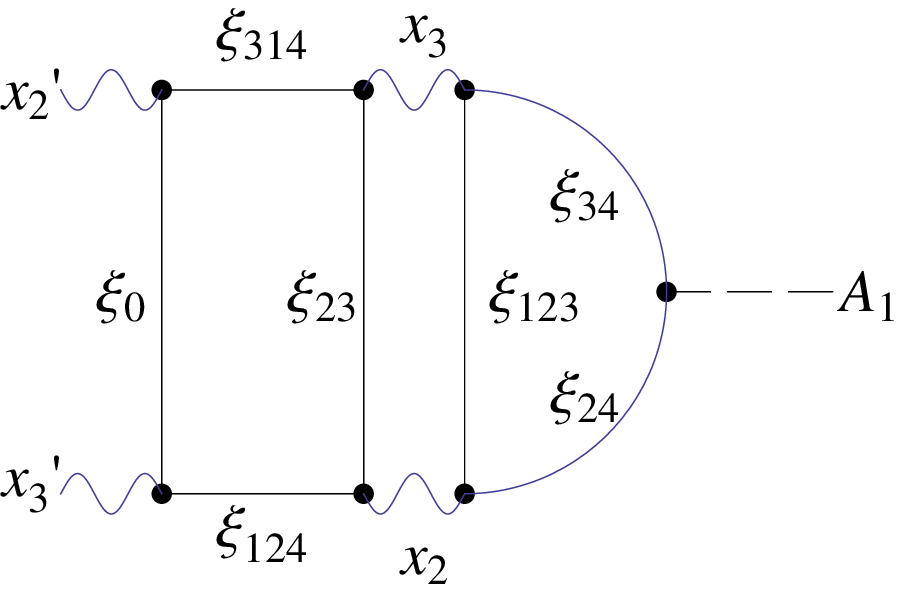}
\caption{The triangle loop diagram of axial anomaly. $A_1 x_2' x_3'$ type and its rescsttering.}
\label{g1d}
\end{center}
\end{minipage}
\end{figure}

\begin{figure}
\begin{minipage}[b]{0.47\linewidth}
\begin{center}
\includegraphics[width=6cm,angle=0,clip]{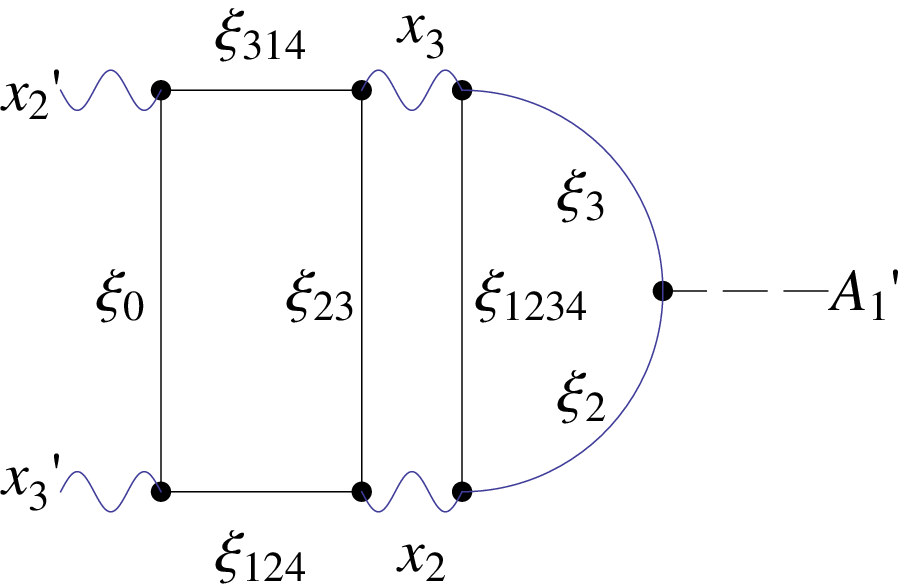}
\caption{The triangle loop diagram of axial anomaly. $A_1 x_3' x_2'$ type and its rescattering.} 
\label{g1c}
\end{center}
\end{minipage}
\hfill
\begin{minipage}[b]{0.47\linewidth}
\begin{center}
\includegraphics[width=6cm,angle=0,clip]{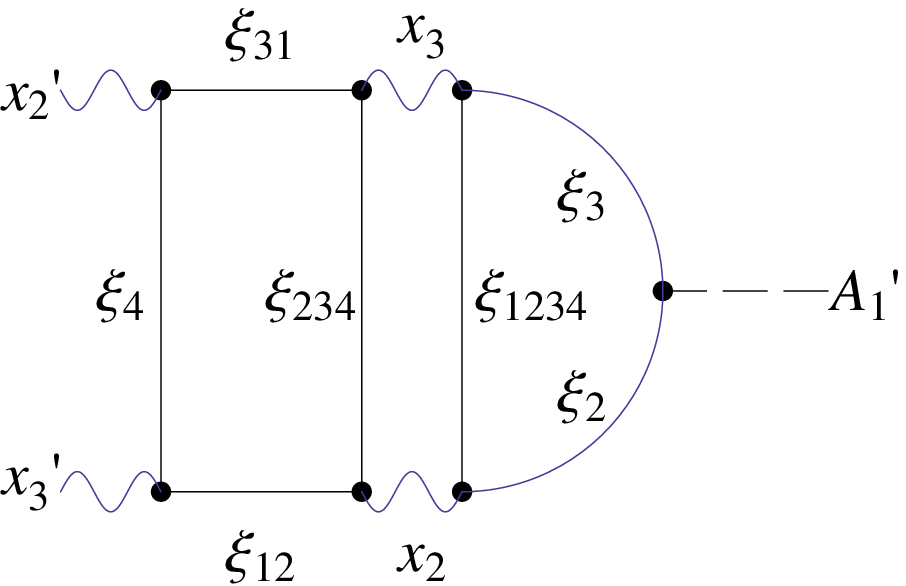}
\caption{The triangle loop diagram of axial anomaly. $A_1 x_2' x_3'$ type and its rescsttering.}
\label{g1d}
\end{center}
\end{minipage}
\end{figure}

\begin{figure}
\begin{minipage}[b]{0.47\linewidth}
\begin{center}
\includegraphics[width=6cm,angle=0,clip]{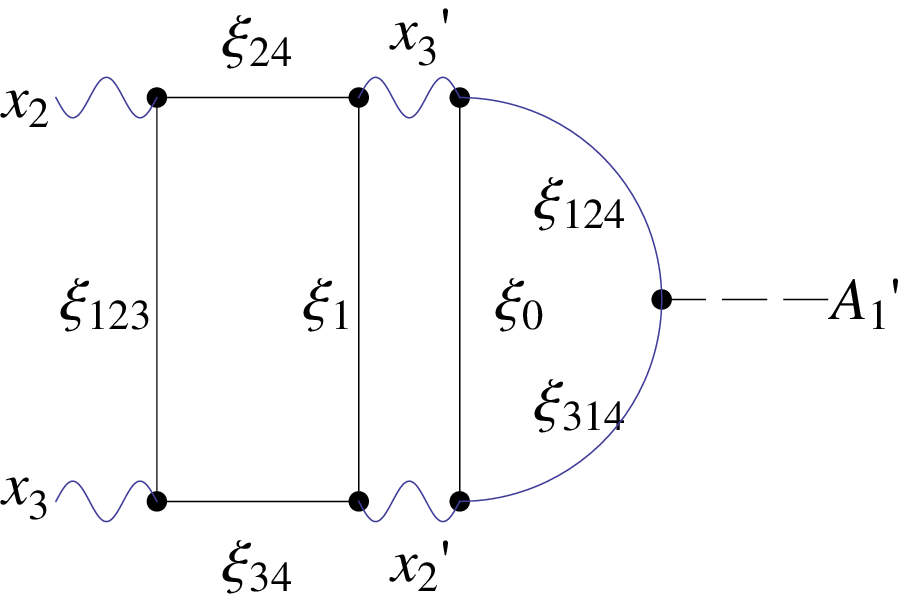}
\caption{The triangle loop diagram of axial anomaly. $A_1 x_3' x_2'$ type and its rescattering.} 
\label{g1c}
\end{center}
\end{minipage}
\hfill
\begin{minipage}[b]{0.47\linewidth}
\begin{center}
\includegraphics[width=6cm,angle=0,clip]{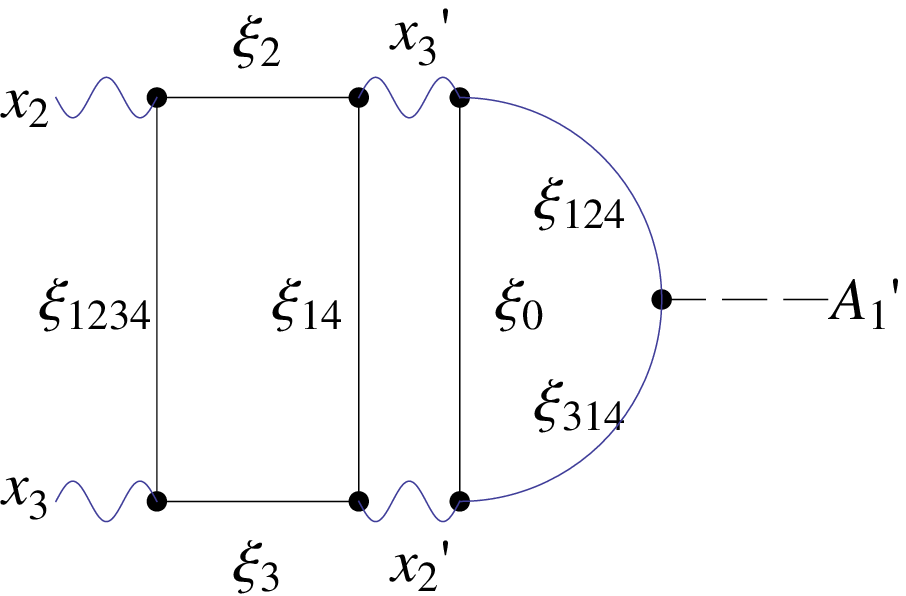}
\caption{The triangle loop diagram of axial anomaly. $A_1 x_2' x_3'$ type and its rescsttering.}
\label{g1d}
\end{center}
\end{minipage}
\end{figure}

\begin{figure}
\begin{minipage}[b]{0.47\linewidth}
\begin{center}
\includegraphics[width=6cm,angle=0,clip]{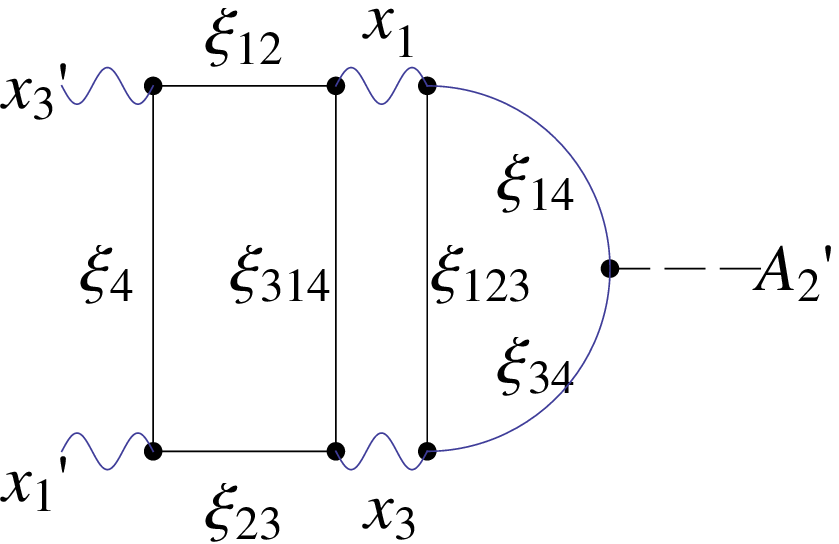}
\caption{The triangle loop diagram of axial anomaly.  $A_2 x_3 x_1$ type and its rescattering.}
\label{g2}
\end{center}
\end{minipage}
\hfill
\begin{minipage}[b]{0.47\linewidth}
\begin{center}
\includegraphics[width=6cm,angle=0,clip]{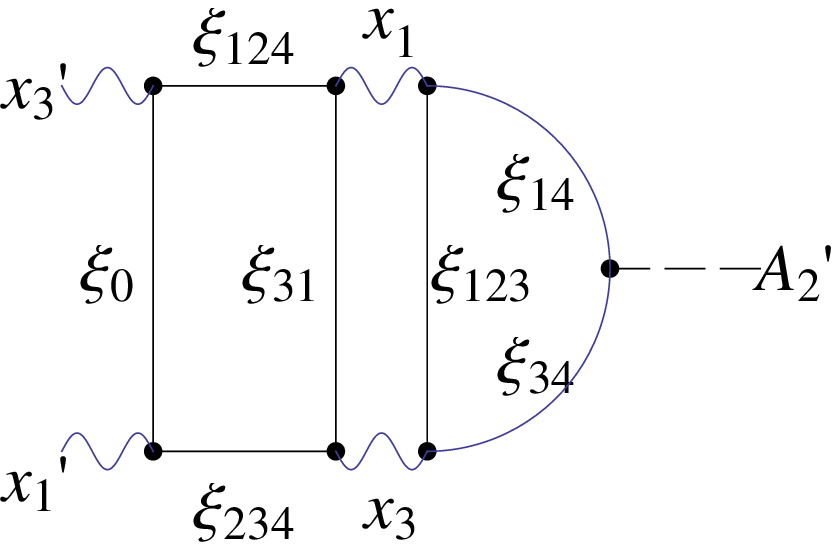}
\caption{The triangle loop diagram of axial anomaly.  $A_2 x_3 x_1$ type and its rescattering.}
\label{g2d}
\end{center}
\end{minipage}
\end{figure}

\begin{figure}
\begin{minipage}[b]{0.47\linewidth}
\begin{center}
\includegraphics[width=6cm,angle=0,clip]{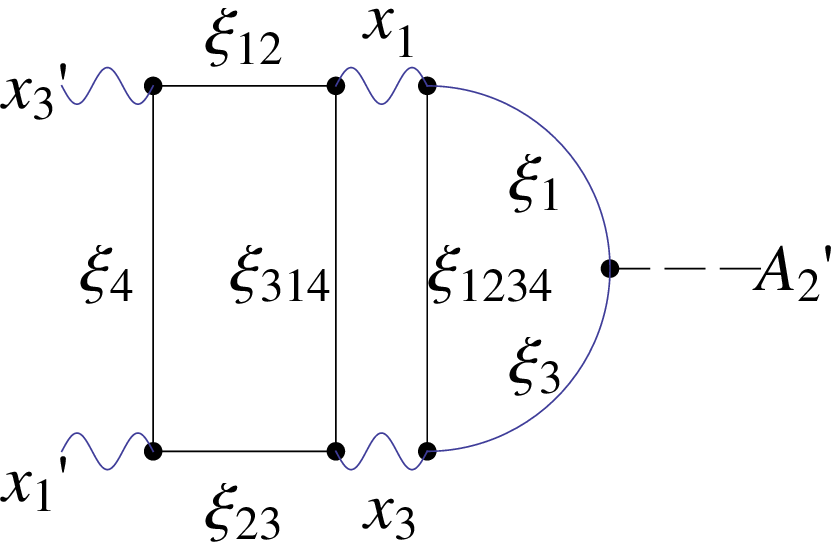}
\caption{The triangle loop diagram of axial anomaly.  $A_2' x_3 x_1$ type and its rescattering.} 
\label{g2e}
\end{center}
\end{minipage}
\hfill
\begin{minipage}[b]{0.47\linewidth}
\begin{center}
\includegraphics[width=6cm,angle=0,clip]{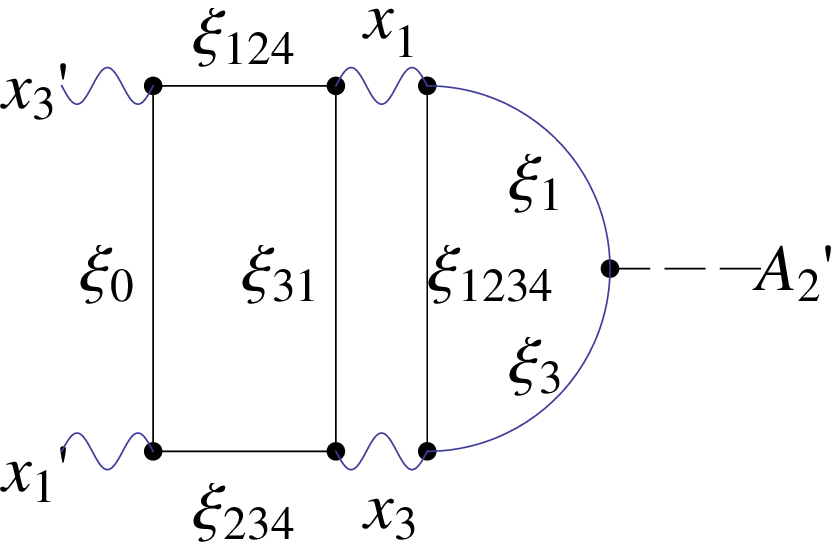}
\caption{The triangle loop diagram of axial anomaly.  $A_2' x_3 x_1$type and its rescattering.} 
\label{g2f}
\end{center}
\end{minipage}
\end{figure}

\begin{figure}
\begin{minipage}[b]{0.47\linewidth}
\begin{center}
\includegraphics[width=6cm,angle=0,clip]{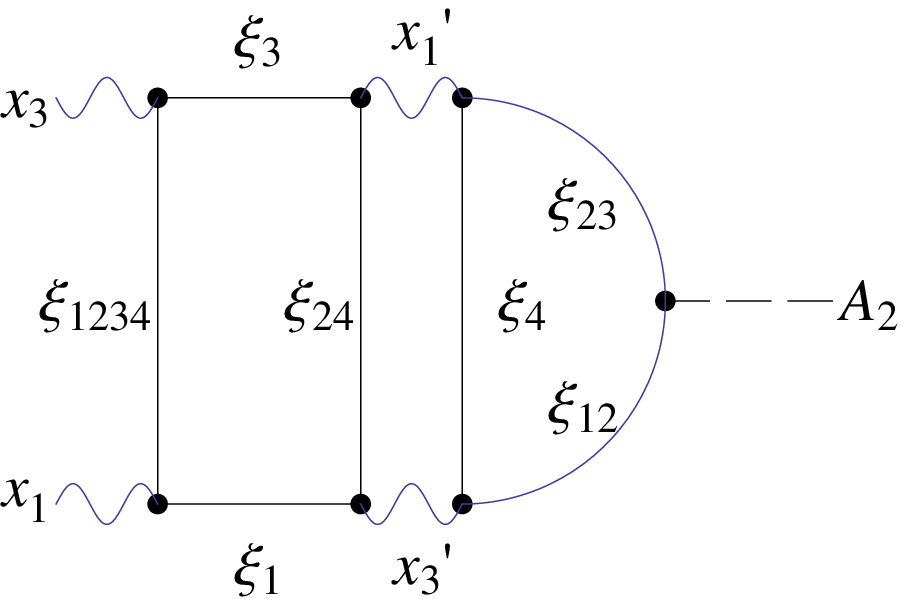}
\caption{The triangle loop diagram of axial anomaly.  $A_2 x_3 x_1$ type and its rescattering.}
\label{g2}
\end{center}
\end{minipage}
\hfill
\begin{minipage}[b]{0.47\linewidth}
\begin{center}
\includegraphics[width=6cm,angle=0,clip]{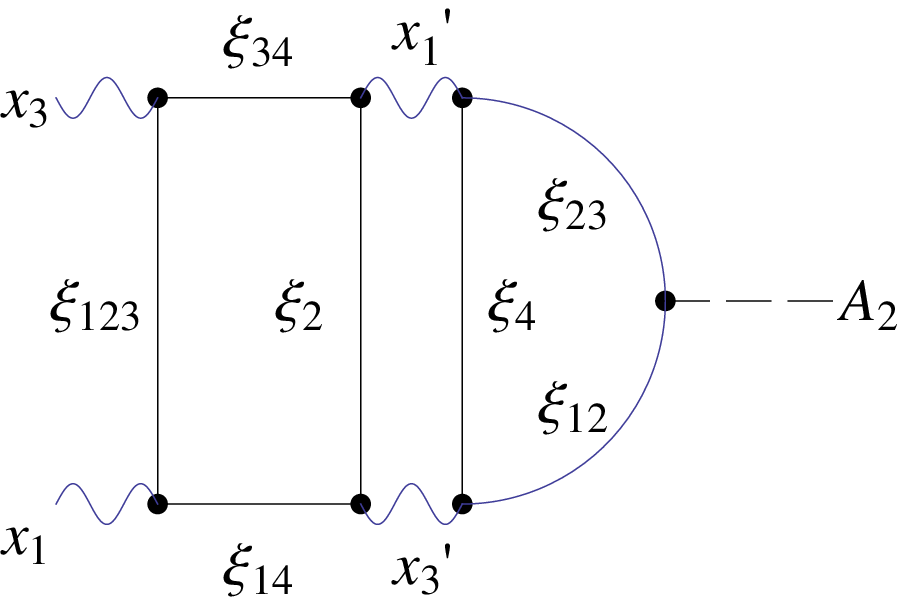}
\caption{The triangle loop diagram of axial anomaly.  $A_2 x_3 x_1$ type and its rescattering.}
\label{g2d}
\end{center}
\end{minipage}
\end{figure}

\begin{figure}
\begin{minipage}[b]{0.47\linewidth}
\begin{center}
\includegraphics[width=6cm,angle=0,clip]{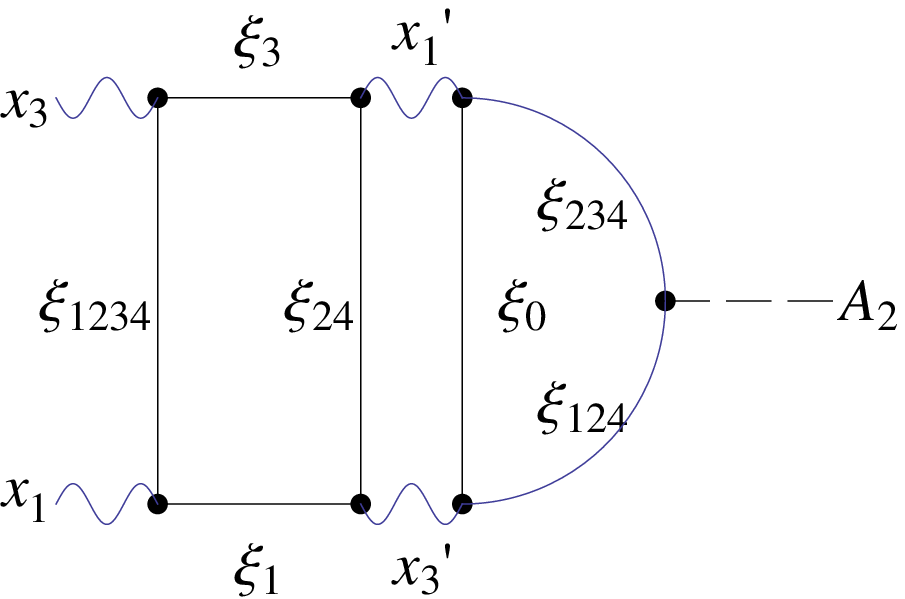}
\caption{The triangle loop diagram of axial anomaly.  $A_2' x_3 x_1$ type and its rescattering.} 
\label{g2e}
\end{center}
\end{minipage}
\hfill
\begin{minipage}[b]{0.47\linewidth}
\begin{center}
\includegraphics[width=6cm,angle=0,clip]{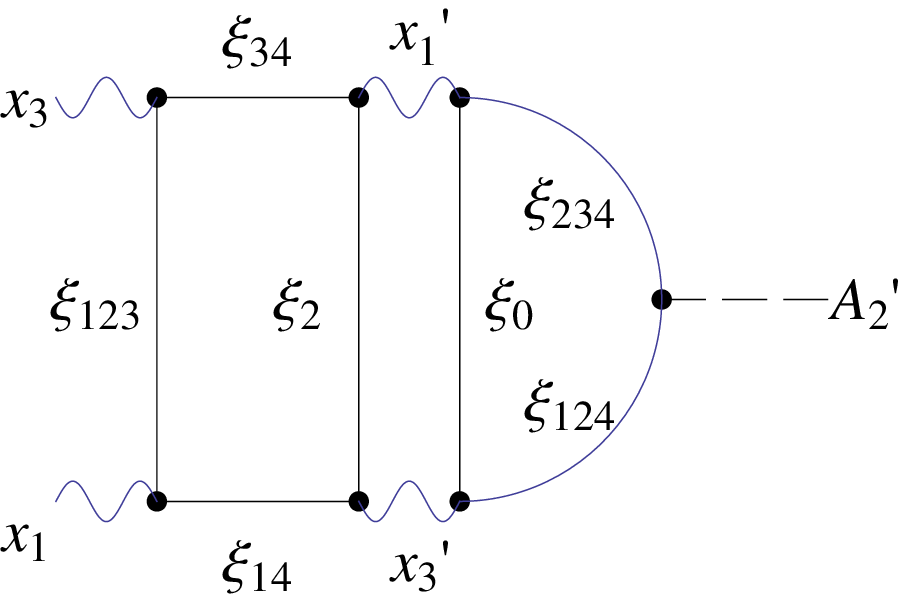}
\caption{The triangle loop diagram of axial anomaly.  $A_2' x_3 x_1$type and its rescattering.} 
\label{g2f}
\end{center}
\end{minipage}
\end{figure}

\begin{figure}
\begin{minipage}[b]{0.47\linewidth}
\begin{center}
\includegraphics[width=6cm,angle=0,clip]{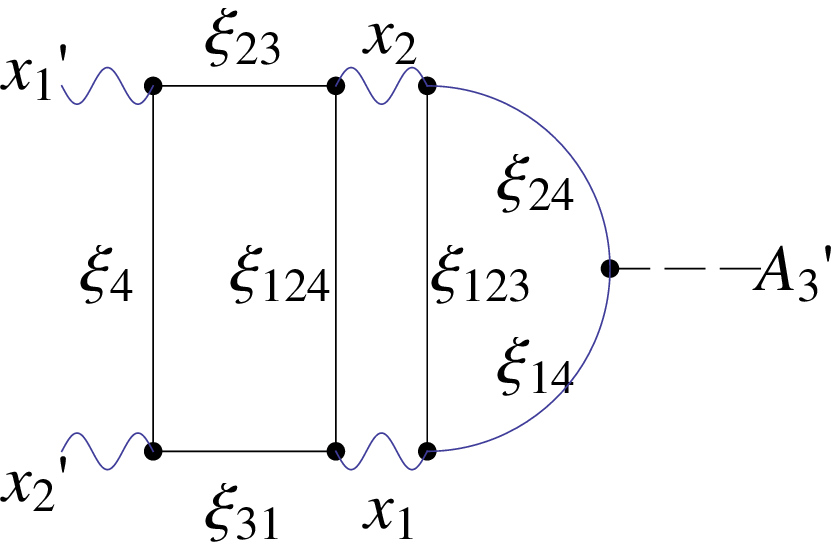}
\caption{The triangle loop diagram of axial anomaly.  $A_3' x_1 x_2$ type and its rescattering.}
\label{g3b}
\end{center}
\end{minipage}
\hfill
\begin{minipage}[b]{0.47\linewidth}
\begin{center}
\includegraphics[width=6cm,angle=0,clip]{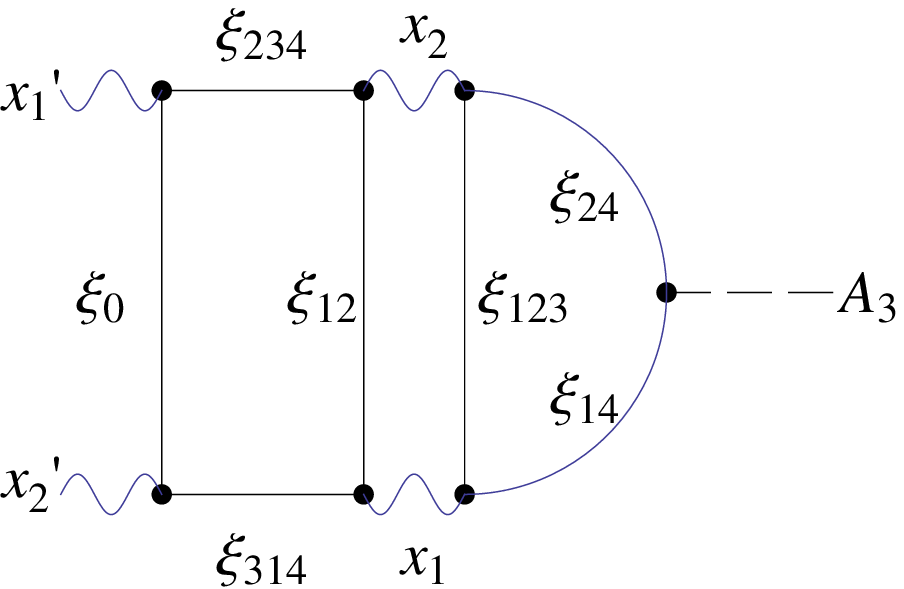}
\caption{The triangle loop diagram of axial anomaly.  $A_3' x_1 x_2$ type and its rescattering.} 
\label{g3c}
\end{center}
\end{minipage}
\end{figure}
\begin{figure}
\begin{minipage}[b]{0.47\linewidth}
\begin{center}
\includegraphics[width=6cm,angle=0,clip]{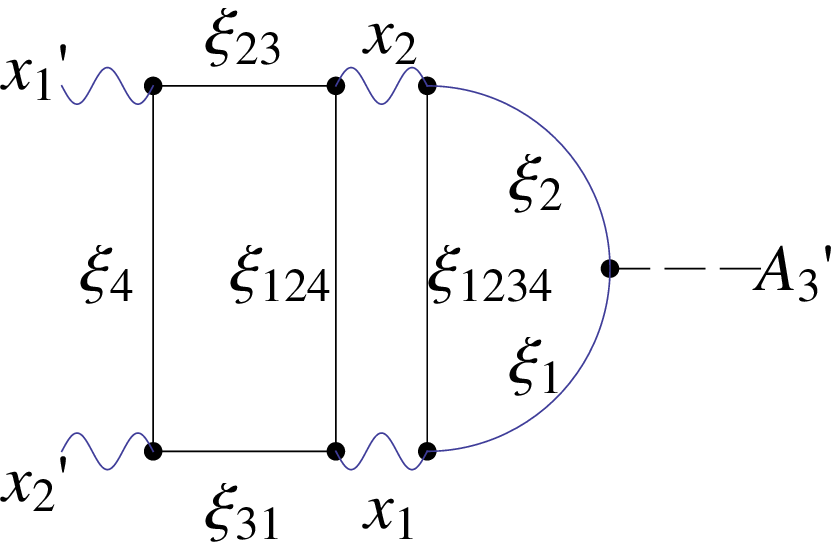}
\caption{The triangle loop diagram of axial anomaly.  $A_3' x_1 x_2$ type and its rescattering.} 
\label{g3e}
\end{center}
\end{minipage}
\hfill
\begin{minipage}[b]{0.47\linewidth}
\begin{center}
\includegraphics[width=6cm,angle=0,clip]{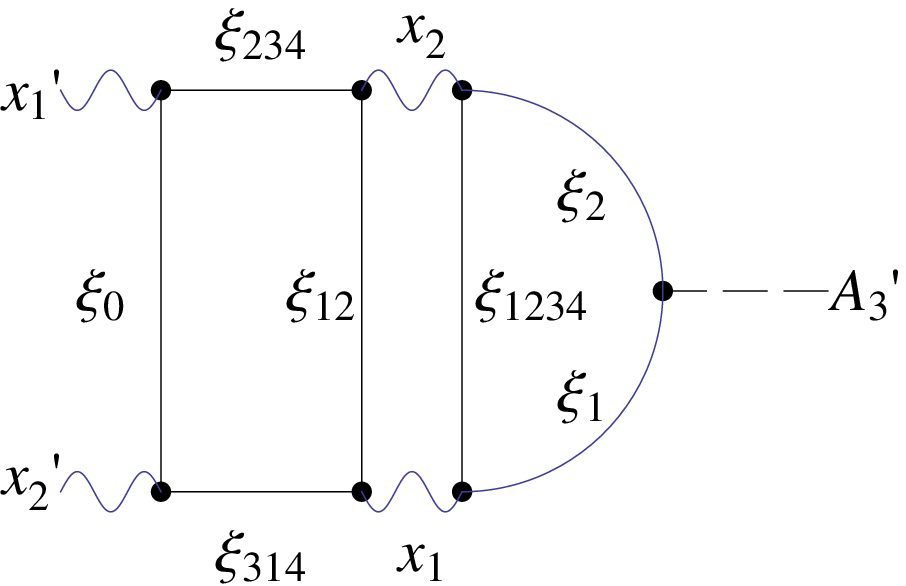}
\caption{The triangle loop diagram of axial anomaly.  $A_3' x_1' x_2'$ type and its rescattering.}
\label{g3f}
\end{center}
\end{minipage}
\end{figure}

\begin{figure}
\begin{minipage}[b]{0.47\linewidth}
\begin{center}
\includegraphics[width=6cm,angle=0,clip]{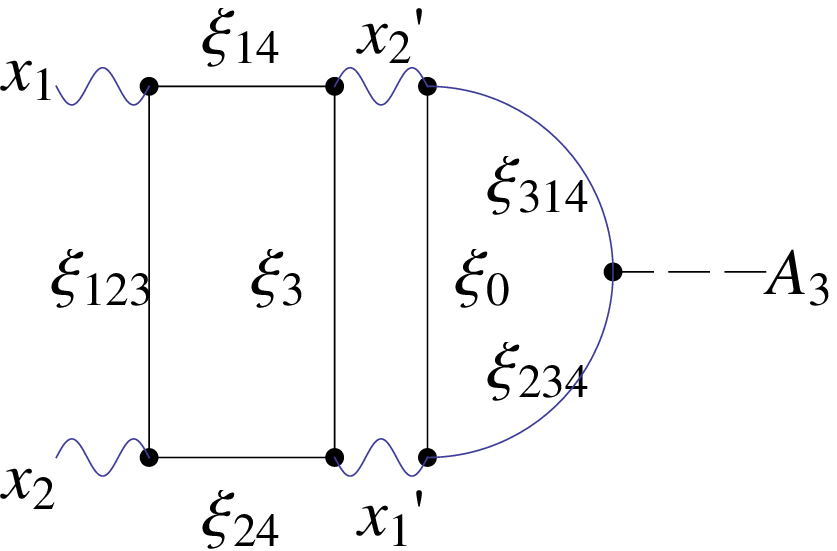}
\caption{The triangle loop diagram of axial anomaly.  $A_3 x_1' x_2'$ type and its rescattering.} 
\label{g3e}
\end{center}
\end{minipage}
\hfill
\begin{minipage}[b]{0.47\linewidth}
\begin{center}
\includegraphics[width=6cm,angle=0,clip]{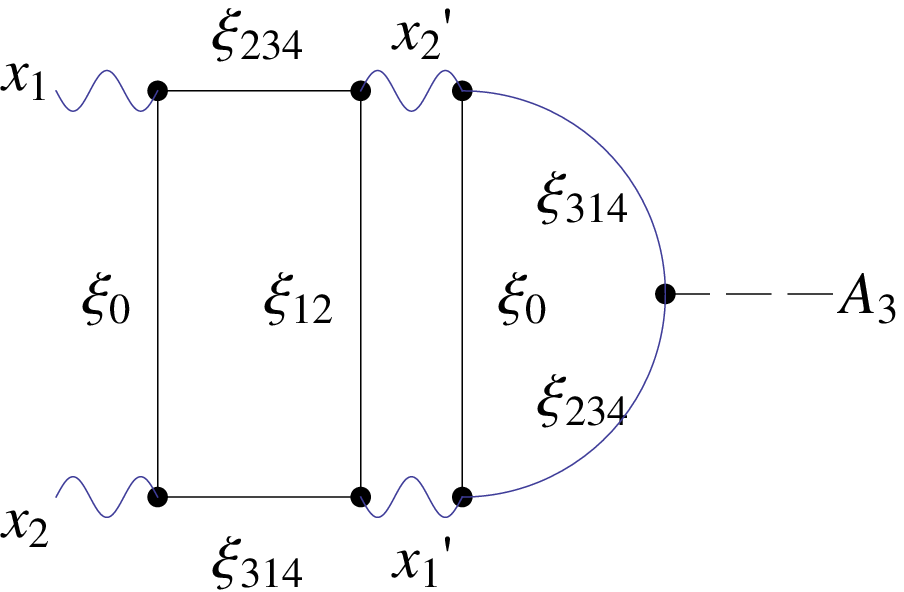}
\caption{The triangle loop diagram of axial anomaly.  $A_3 x_1' x_2'$ type and its rescattering.}
\label{g3f}
\end{center}
\end{minipage}
\end{figure}

\begin{figure}
\begin{minipage}[b]{0.47\linewidth}
\begin{center}
\includegraphics[width=6cm,angle=0,clip]{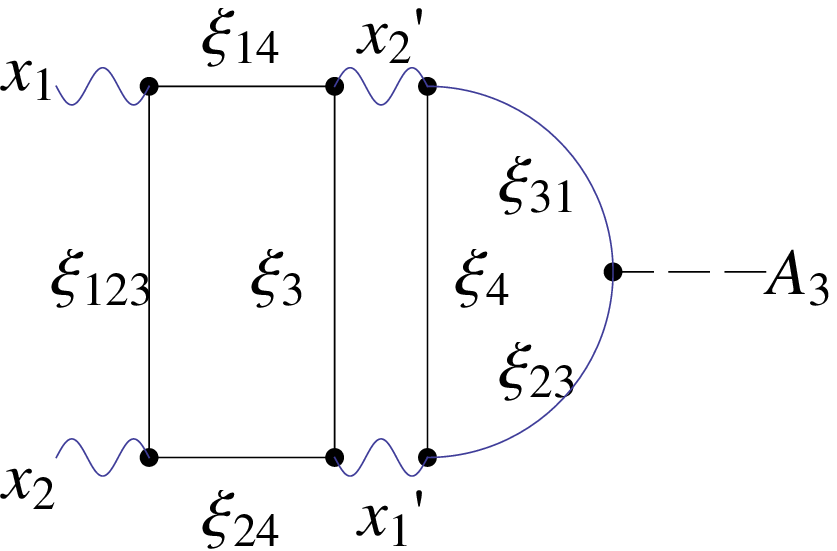}
\caption{The triangle loop diagram of axial anomaly.  $A_3 x_1' x_2'$ type and its rescattering.} 
\label{g3e}
\end{center}
\end{minipage}
\hfill
\begin{minipage}[b]{0.47\linewidth}
\begin{center}
\includegraphics[width=6cm,angle=0,clip]{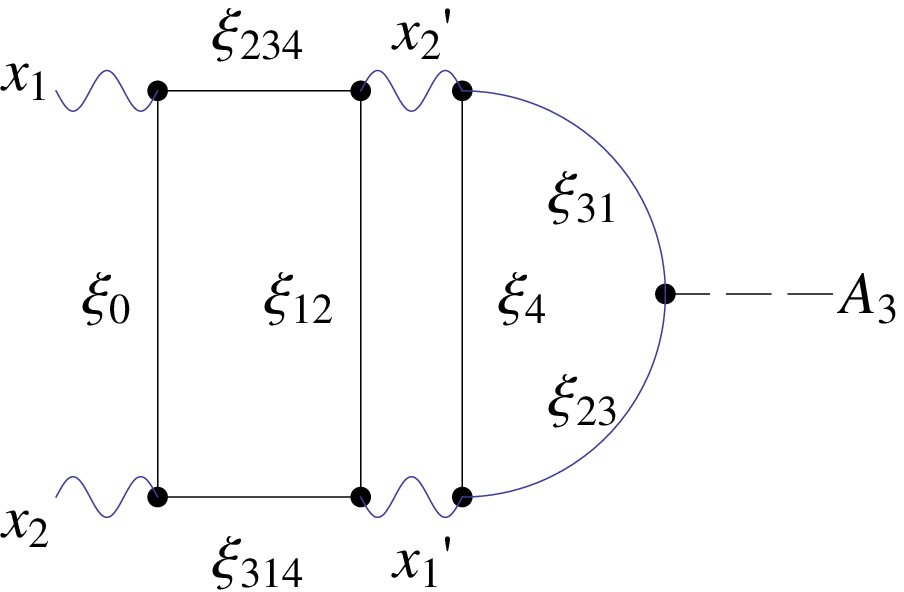}
\caption{The triangle loop diagram of axial anomaly.  $A_3 x_1' x_2'$ type and its rescattering.}
\label{g3f}
\end{center}
\end{minipage}
\end{figure}
 The Dirac spinor interacts with self-dual vector fields $x_1, x_2, x_3$ and axial current vertices $A_1,A_2,A_3$. We consider instanton contribution, and select the momentum transfer at the axial vector vertex to be 0.

We define the eigenfunctions
\begin{equation}
H_0^{\pm}{\psi^\pm}_{p_i}(p)=\nu_p {\psi^\pm}_{p_i}(p),\qquad H_0^{\pm}{\phi^\pm}_{p_i}(p)=-\nu_p {\phi^\pm}_{p_i}(p),
\end{equation}
where $\nu_p=\sqrt{|p|^2+m^2}$ positive, and 
\begin{equation}
\psi^+_{p_i}(p)=\frac{1}{N(p)}\left(\begin{array}{c} p_i\\
                                                             \nu_p+m\end{array}\right),\qquad
\psi^-_{p_i}(p)=\frac{1}{N(p)}\left(\begin{array}{c} \nu_p+m\\
                                                              p_i\end{array}\right) 
\end{equation}
and
\begin{equation}
\phi^+_{p_i}(p)=\frac{1}{N(p)}\left(\begin{array}{c} -(\nu_p+m) \\
                                                             p_i\end{array}\right) ,\qquad
\phi^-_{p_i}(p)=\frac{1}{N(p)}\left(\begin{array}{c} -p_i \\
                                                              \nu_p+m\end{array}\right).
\end{equation}

The normlization is $N(p)=\sqrt{2\nu_p(\nu_p+m)}$, and the overlaps of the wave functions are
\begin{equation}
[\psi^-_{p_i}]^\dagger \psi^+_{q_j}=[\phi^+_{p_i}]^\dagger \phi^-_{q_j}=\frac{p_i}{\nu_p}=P_{pq}
\end{equation}
and 
\begin{equation}
[\phi^-_{p_i}]^\dagger \psi^+_{q_j}=-[\psi^+_{p_i}]^\dagger \phi^-_{q_j}=\frac{m}{\nu_p}\delta_{pq}=Q_{pq}
\end{equation}

We remove the suffices $\pm$, and define $\phi^-_p=c\phi_p$ and $\psi^-_p=c\psi_p$.
The combination $(\psi_p, c\psi_p)$ and $(\phi_p, c\phi_p)$ are Dirac spinors of a particle
and an anti particle, respectively.

The triangle diagram that introduces the axial anomaly is produced by taking the propagator between the two vector vertices $x_i,x_j$ to be $\xi_0, \xi_{1234}$, $\xi_4$ or $\xi_{123}$. In the case of $A_1 x_2 x_3$ vertex,  to incorporate antisymmetry of $x_2$ and $x_3$, I define the propagator as $\displaystyle \frac{-\gamma_1 p_1+m}{p_1^2+m^2}\gamma_0$.
The Figs.2-5 give the anomaly contribution,
\begin{eqnarray}
\Gamma_{A}&=&Tr[ (\gamma_3 x_3)\frac{\gamma_1p_1+\gamma_3  Q_3+m}{p_1^2+Q_3^2+m^2}
( \gamma_1\gamma_5  A_1)\frac{-\gamma_2 Q_2-\gamma_1 p_1+m}{p_1^2+Q_2^2+m^2}\nonumber\\
&&\times( \gamma_2  x_2)\frac{-\gamma_1 p_1+m}{p_1^2+m^2}\gamma_0]
\end{eqnarray}

The trace becomes
\begin{eqnarray}
&&-4i\, m (p_1^2+m^2) A_1 x_2 x_3 \frac{ 1 } {(p_1^2+Q_3^2+m^2)(p_1^2+Q_2^2+m^2)(p_1^2+m^2)}\nonumber\\
&&=-4i\, m  A_1 x_2 x_3 \frac{1} {(p_1^2+Q_3^2+m^2)(p_1^2+Q_2^2+m^2)}.
\end{eqnarray}
The imaginary part will correspond to the instanton contribution. 
The Figs.6-9 give
\begin{eqnarray}
\Gamma_{B}&=& Tr[ ( \gamma_1  x_1) \frac{\gamma_2 p_2+\gamma_1 Q_1+m}{p_2^2+Q_1^2+m^2}
(\gamma_2 \gamma_5  A_2) \frac{-\gamma_2 p_2+\gamma_2 Q_3+m}{p_2^2+Q_3^2+m^2}\nonumber\\
&&\times( \gamma_3  x_3) \frac{-\gamma_2 p_2+m}{p_2^2+m^2}\gamma_0]
\end{eqnarray}

The trace becomes
\begin{eqnarray}
&&-4i\, m (p_2^2+m^2) x_1 A_2 x_3 \frac{1} {(p_2^2+Q_1^2+m^2)(p_2^2+Q_3^2+m^2)(p_2^2+m^2)}\nonumber\\
&&=-4i\, m  x_1 A_2 x_3 \frac{1} {(p_2^2+Q_1^2+m^2)(p_2^2+Q_3^2+m^2)}
\end{eqnarray}

The Figs.10-13 give
\begin{eqnarray}
\Gamma_{C}&=&Tr[ ( \gamma_2  A_2) \frac{\gamma_3 p_3+\gamma_2 Q_2+m}{p_1^2+Q_2^2+m^2}
(\gamma_3\gamma_5  A_3)\frac{\gamma_3 p_3+\gamma_1  Q_1+m}{p_3^2+Q_1^2+m^2}\nonumber\\
&&\times (\gamma_1  A_1) \frac{-\gamma_3 p_3+m}{ p_3^2+m^2}\gamma_0]
\end{eqnarray}

The trace becomes 
\begin{eqnarray}
&&-4 i\, m (p_3^2+m^2) x_1 x_2 A_3 \frac{1} {(p_3^2+Q_2^2+m^2)(p_3^2+Q_1^2+m^2)(p_3^2+m^2)}\nonumber\\
&&=-4i\, m x_1 x_2 A_3 \frac{1} {(p_3^2+Q_2^2+m^2)(p_3^2+Q_1^2+m^2)}
\end{eqnarray}

The two gauge particle propagators are $\displaystyle \frac{1}{Q^2}$ and $\displaystyle \frac{1}{{Q'}^2}$.
\begin{eqnarray}
&&\int dQ 4\pi Q^2\int dQ' 4\pi Q'^2 \frac{1} {Q^2 Q'^2 (p^2+Q^2+m^2)(p^2+Q'^2+m^2)}\nonumber\\
&&=(\int dQ\frac{4\pi }{(p^2+Q^2+m^2)})^2=\frac{4\pi^3}{p^2+m^2}
\end{eqnarray}

The integration of the quark momentum $p$ yields the quantity proportional to the free energy density in the finite temperature field theory\cite{Kaj94}.   
We parametrize $p^2+m^2=\omega_n^2+\bar p^2+m^2=\omega_n^2+E^2$ and calculate 
\begin{eqnarray}
I(m)&=&\int \frac{d^3 p}{(2\pi)^3}(\frac{1 }{2E}+\frac{1}{E}\frac{1}{e^{E/T}-1})\nonumber\\
&=&T\int\frac{d^3 p}{(2\pi)^3}\frac{1}{\bar p^2+m^2}+2T\sum_{n=1}^\infty \frac{1}{(2\pi Tn)^2+\bar p^2+m^2}\nonumber\\
&=&T\int\frac{d^3 p}{(2\pi)^3}\frac{1}{\bar p^2+m^2}+2T\sum_{l=0}^\infty\sum_{n=1}^\infty\int \frac{d^3 p}{(2\pi)^3}
\frac{(-m^2)^l}{[(2\pi n T)^2+\bar p^2]^{l+1}}
\end{eqnarray}
where
\[
\int \frac{d^3 p}{(2\pi)^3}\frac{1}{E}\frac{1}{e^{E/T}-1}
\]
is modified to the sum of the contribution from $\displaystyle \frac{E}{T}=2\pi i n$ and the complex integral formula
\[
T\sum_n f(2\pi n T)=\oint \frac{dz}{2\pi}f(z)\frac{1}{e^{i\beta z}-1} 
\]
is used.

The integral can be done in usual dimensional regularization. Taking the zero-temperature part in the dimensional regularization
\[
\int\frac{d^3 p}{(2\pi)^3} \frac{1}{\bar p^2+m^2}=-\frac{m}{4\pi}
\]
one finds
\begin{eqnarray}
T\mu^{2\epsilon}\int\frac{d^{3-2\epsilon}p}{(2\pi)^{3-2\epsilon}}\frac{1}{\bar p^2+m^2}&=&-\frac{mT}{4\pi}+\frac{T^2}{12}-\frac{m^2}{16\pi^2}[\frac{1}{\epsilon}+\log\frac{\bar\mu^2}{T^2}-2(\log 4\pi-\gamma)]\nonumber\\
&&+\frac{\zeta(3)}{128\pi^4}\frac{m^4}{T^2}+\cdots
\end{eqnarray}

We have summed up the contribution from triangle vertices $A_1x_2'x_3'$, $x_1A_2'x_3$, $x_1'x_2'A_3$
and $x_1x_2 A_3'$. The vertices of $x_2'\xi_{31}\xi_4$ and $x_3'\xi_{12}\xi_4$ or $x_2'\xi_{314}\xi_0$ and $x_3'\xi_{124}\xi_0$ have the same sign, and products has the positive sign.  Relative signs of the vertices $A_1\xi_{12}\xi_{31}$ and  $A_1\xi_{124}\xi_{314}$ is not known apriori, but via transformation $G_{23}$, $\xi_{31}$ and $\xi_{314}$ are interchanged and the sign of the vertices $x_1\xi_{12}\xi_{314}$ and $x_1\xi_{124}\xi_{31}$
are opposite.  The sign of vertices  $A_2'\xi_{14}\xi_{34}$ ,  $A_2'\xi_1\xi_3$, $A_3\xi_{314}\xi_{234}$, $A_3 \xi_{31}\xi_{23}$, $A_3'\xi_{24}\xi_{14}$ and $A_3'\xi_2\xi_1$, which we denote $A_i \xi_{even/odd}\xi_{even/odd}$ can be defined from those of $x_i \xi_{even/odd} (G_{23}\xi_{even/odd})=x_i\xi_{even/odd}\xi_{odd/even}$, which are fixed by the tri-linear form ${^t\phi}CX\psi$. Hence in the Coulomb gauge, the vertices  $A_i\xi_{even/odd}\xi_{even/odd}$ and $A_i'\xi_{even/odd}\xi_{even/odd}$ for $i=1,2,3$ can be defined by assuming invariance under $G_{23}$.

 In the $\pi^0 \to\gamma\gamma$ decay the model based on eq.(\ref{divergence}) was successful but in the $\eta\to\gamma\gamma$ or $\eta'\to\gamma\gamma$ there appears ambiguities in the relative strength of $j_\mu^{53}$ and $j_\mu^{58}$\cite{Ioffe08}, in which instantons could play an essential role.
t'Hooft's hypothesis that singularities of the amplitudes calculated in QCD on the quark-gluon basis reproduces that on the hadron basis was questioned in \cite{Ioffe08}. In our model the vector particles exchanged in the intermediate state after the triangle vertex are photons, but in the infrared, gluon exchange admixture would become important\cite{Ioffe08}.
Polarization of the intermediate gluon depends on the momentum direction of the initial meson in non-trivial way, and whether the singularity in the quark-gluon level does not reproduce that of the hadron level is not so evident.      

\section{Summary and outlook}

We studied the axial vector current coupling to quark triangle diagram defined by the divergence of the axial vector current, using the triality symmetry of quarks expressed by octonions.  The axial anomaly appears as a reflection of the symmetry of the matter field. 
In the PCAC theory, a pion field is described as a divergence of the axial current
\[
\partial_\mu A^{\mu a}=f_\pi m_\pi^2\phi^a.
\]    
Since the axial current anomaly appears as a reflection of the triality symmetry of matter field, one could argue that condensates whose quantities have commonly viewed as constant empirical mass-scales that fill all-space time, are instead wholly contained within hadrons\cite{BS08, BRST12}. 

It is argued in \cite{RW12} that the QCD condensates are vacuum properties and not hadron properties. In our opinion, the vacuum in which vector fields are produced is defined by the triality symmetry of spinors that interact with the vector fields, and the vacuum properties are in fact defined by hadrons when we observe vector fields.  As discussed in \cite{BRST12}, the magnetic field inside the super conductor is expelled by the super current induced by matter and the Cooper pair condensates are constants only within the matter that supports them.

The triality symmetry could explain the discrepancy of the critical flavor number of QCD in the SF scheme and in the MOM scheme.
The momentum dependence of the running coupling $\alpha_s(q)$ derived in MOM scheme lattice simulation of domain wall fermions, suggests that the system of $N_f=2+1$ is not so far from the conformal window\cite{FN08}. 
 In the SF, matching of lattice configurations of different scales on spacial boundaries is necessary, and one could attribute the extra flavor number in SF as compared to the MOM scheme as due to the triality symmetry
which could be selected in the MOM scheme. 

 If $e^-, \mu^-$ and $\tau^-$ belong to their definite triality sector,  EM interaction is sensitive to the triality, and quarks in the strong interaction and neutrinos in the weak interaction do not select the triality, one can produce neutrino mass without invoking right-handed neutrinos\cite{SF12}.

Leptons in the detector of our telescope belongs to a quaternion basis. Photons emitted from by the EM interactions in the triality sector different from that of our telescope will not be detected. In the space of octonions, except the instanton contribution that appears in the world transformed by $G_{23}$, galaxies transformed by five transformations $G_{23}, G_{12}, G_{13}, G_{123}$ and $G_{132}$ will appear as non-luminous matter, or a dark matter. 

 Astronomically, measurements with WMAP space crafts confirm that almost exactly 5 times more dark matter than the normal matter are observed\cite{Bl11}, which coincides with the number of triality symmetry transformations of an octonion.
 The spectral analysis of 511keV emission from positron annihilation from the galaxy\cite{Prantzos11} also 
suggests that the cluster of dark matter, which could be particles in a different triality sector than
that in our solar system are distributed in the galaxy. 

The cosmological model with 'standard' Einstein's Cosmology Constant $\Lambda$ defines the density $\displaystyle \rho_\Lambda=\frac{\Lambda c^2}{8\pi G}$,  and the Hubble expansion rate $H$ defines the density $\displaystyle \rho_{crit}=\frac{3H^2}{8\pi G}$. The ratio of the dark matter density $\rho_{DM}$ devided by the critical density $\rho_{crit}$ i.e. $\displaystyle \frac{\rho_{DM}}{\rho_{crit}}=\Omega_{DM}$ is about 24\% while the ratio of baryon density $\displaystyle \frac{\rho_{baryon}}{\rho_{crit}}=\Omega_{baryon}$ is about 4\%, which means that the ratio of the mass of the dark matter and the normal matter in the universe as a whole is almost 5 as expected.

\newpage
{\bf Acknowledgement}\\
The author thanks Stan Brodsky for attracting attention to the refs.\cite{Ioffe08,BS08,BRST12} and sending encouragements. Helpful lectures by Prof. Kajantie at Inst. of Theor. Phys. Uni. Helsinki in February 1994 are also acknowledged.  


\begin{thebibliography}{99}
\bibitem{BLM83}  Blodsky,S.J., Lepage, G.P. and MacKenzie,P.B.: On the elimination of scale ambiguities in perturbative quantum chromodynamics, \PRD{\bf 28} 228 (1983).
\bibitem{ZWZ83} Zumino, B. ,Wu, Yong-Shi  and Zee,A.: Chiral Anomalies, Higher Dimensions and   
Differential Geometry, \NPB{\bf 239}, 477 (1983).
\bibitem{Gr89} Grunberg,G.: On some ambiguities in the method of effective charge, \PRD{\bf 40} 680 (1989).
\bibitem{GK98} Gardi,E. and Karliner,M. : Relations between Observables and the Infrared Fixed-Point in QCD, \NPB{\bf 529}383(1998) , arXiv:98022184[hep-ph]
\bibitem{GGK98} Gardi,E., Grunberg, G. and Karliner, M.: Can the QCD running coupling have a causal analytical structure, JHEP 9807:007 (1998)
\bibitem{Gr02} Grunberg,G.: Fixing the Conformal Window in QCD, \PRD{\bf 65} 021701(R) (2002)
\bibitem{BZ82} Banks,T. and Zaks, A.: On the phase structure of vector-like gauge theories with massless fermions, \NPB {\bf 196} 189 (1982) 
\bibitem{AFN08} Appelquist,T. , Fleming. G.T. and Neil E.T.,: Lattice Study of the Conformal Window in QCD-like Theories, \PRL{\bf 100},171607(2008), Errata \PRL{\bf 102},149902(2009).

\bibitem{CHPS13} Cheng, A. ,Hasenfratz, A., Petropoulos, G. and Schaich, D. : Scale-dependent mass anomalous dimension from Dirac eigenmodes, arXiv:1301.1355[hep-lat].
\bibitem{BRRQ13} Bashir, A. , Reya, A. and Rodriguez-Quintero, J. : QCD:Restoration of Chiral Symmetry and Deconfinement for Large $N_f$, arXiv:1302.5829 [hep-lat].
\bibitem{SCHP12} Schaich, D., Cheng, A. Hasenfratz,A. and Petropoulos, G. :Bulk and finite-temperature transitions in SU(3) gauge theories with many light fermions, Lattice12 Proc. arXiv:1207.7164[hep-lat]
\bibitem{ABBCRQ12} Ayala,A. , Bashir, A. , Binosi, D. ,Christoforetti, M. and Rodriguez-Quintero,J. : Quark flavor effects on gluon and ghost proppagators, \PRD {\bf 86}, 074512(2012).

\bibitem{BGGR01} Brodsky,S.J., Gardi.E., Grunberg,G. and Rathdan,J. : Disentangling running coupling and conformal effects in QCD, \PRD {\bf 63}, 094017 (2001)
\bibitem{FN08} Furui,S. and Nakajima,H., Roles of the quark field in the infrared lattice Coulomb gauge and Landau gauge QCD, PoS Lattice 2007, (2008)
\bibitem{BTD10}  Brodsky,S.J. ,  de T\'eramond,G. and  Deur,A.: Nonperturbative QCD Coupling and its $\beta$ function from Light-Front Holography, \PRD {\bf 81}, 096010 (2010), arXiv:1002.3948[hep-ph].
\bibitem{SF11} Furui,S.: Fermion flavors in quaternion basis and infrared QCD,  Few Body Syst. {\bf 52}(2012), arXiv:1104.1225 [hep-ph].
\bibitem{SF12} Furui,S.: The flavor symmetry in the standard model and the triality symmetry, Int J. Mod. Phys. {\bf A27} 1250158(2012)
\bibitem{Wilczek11} Wilczek,F.: BCS as Foundation of Inspiration: The Transmutation of Symmetry, {\it BCS 50 years} ed. by L.N. Cooper and D. Feldman, World Scientific, (2011).
\bibitem{MMPPSS08} Melchiorri et al., Sterile Neutrinos in Light of Recent Cosmological and Oscillaion Data: a Multi-Flavor Scheme Approach, JCAP, 0901:036 (2009)
\bibitem{YK08} Y\"uksel, H. and Kistler, M.D.: Circumscribing late dark matter decays model-independently,
\PRD{\bf 78},023507 (2008), arXiv:0711.2906[astro-ph].
\bibitem{CS08} Cembranos, J.A. and Strigari, L.E.: Diffuse MeV gamma rays and galactic 511keV line from decaying WIMP dark matter, \PRD{\bf 77},123519 (2008), arXiv:0801.0630[astro-ph].
\bibitem{SF09} Furui, S.: Chiral Symmetry and BRST Symmetry Breaking, Quaternion Reality and Lattice Simulation, {\it Strong Coupling Gauge Theory in LHC Era},World Scientific, p.398-400(2011). 
\bibitem{SF10}  Furui, S.: Domain Wall Fermion Lattice Simulation in Quaternion Basis,  {\it The IX international Conference on Quark Confinement and the Hadron Spectrum-QCHS IX}, ed by Llanes-Estrada and Pela\'ez, AIP Conference Proceedigs 1343, p.533(2011), arXiv:0912.5397[hep-lat]
\bibitem{Cartan66}  Cartan,\'E. {\it The theory of Spinors}, Dover Pub. (1966) p.118.
\bibitem{AI89} Ansel'm, A.A. and Iogansen, A.A.: Radiative corrections to the axial anomaly, Pis'ma Zh. Exp. Theor. Fiz. {\bf 49}, 185 (1989).
\bibitem{Adler04} Adler,S.: Anomalies to all orders, arXiv:0405040[hep-th].
\bibitem{Ioffe08} Ioffe, B.L.: Axial anomaly in quantum electro- and chromodynamics and the structure of the vacuum in quantum chromodynamics, Usp. Fiz. Nauk{\bf  178}, 647 (2008), arXiv:0809.0212v2.
\bibitem{BZ99} Baulieu,L. and Zwanziger,D.: Renormalized Non-Covariant Gauges and Coulomb Gauge Limit, \NPB{\bf 548}, 527 (1999)
\bibitem{BZ07} Baulieu,L. and Zwanziger,D.: Renormalized Coulomb Gauge, Braz. J. Phys. {\bf 37}, 293(2007).

\bibitem{Ba69} Bardeen, W.A.: Anomalous Ward Identities in Spinor Field Theories, \PRD{\bf 184},
1848 (1969)
\bibitem{WZ71} Wess,J. and  Zumino,B.: Consequences of anomalous Ward identities, \PLB{\bf 37} , 95 (1971)
\bibitem{Lou00} Lounesto,P., {\it Clifford Algebras and Spinors} 2nd ed.  Cambridge Univ. Press, (2000).
\bibitem{CL84} Cheng,T-P.  and Li, L-F.  {\it Gauge theory of elementary particle physics}, Oxford Science Pub. (1984) p.402. 
\bibitem{Kaj94} Kajantie,K. : Lecture Notes, University of Helsinki (1994)
\bibitem{BS08} Brodsky,S.J. and Shrock, R. :Standard-Model Condensates and the Cosmological Constant, \PLB{\bf 666} , 95 (2008) , arXiv:0803.2554[hep-th].
\bibitem{BRST12} Brodsky,S.J. Roberts, C.D. , Shrock, R. and Tandy, C.: Confinement contains condensates, \PRC{\bf 85}, 065202 (2012), arXiv:1202.2376[nucl-th].
\bibitem{RW12} Reinhardt, H. and Weigel, H. :The vacuum nature of the DCD condensates, arXiv:1201.3262 v2[hep-ph].
\bibitem{Bl11} Blitz,l. :Dark Matter, {\it Scientific American}, Oct. 2011, p.38.
\bibitem{Prantzos11}Prantzos,N. et al., The 511keV emission from positron annihilation in the Galaxy, \RMP {\bf 83}, 1001 (2011), arXiv:1009.4620 [astro-ph.HE].
\end{thebibliography}
\end{document}